\documentclass{eas}
\usepackage{graphicx}
\begin{document}

\title{Interacting winds in massive binaries} 
\author{G.\ Rauw}\address{Institut d'Astrophysique \& de G\'eophysique, Universit\'e de Li\`ege, Belgium}
\begin{abstract}
Massive stars feature highly energetic stellar winds that interact whenever two such stars are bound in a binary system. The signatures of these interactions are nowadays found over a wide range of wavelengths, including the radio domain, the optical band, as well as X-rays and even $\gamma$-rays. A proper understanding of these effects is thus important to derive the fundamental parameters of the components of massive binaries from spectroscopic and photometric observations. 
\end{abstract}
\maketitle
\section{Introduction}
Whenever two massive stars (spectral types O or Wolf-Rayet) form a physical binary system, their stellar winds interact. Historically, this concept was first introduced by Prilutskii \& Usov (\cite{PU}) and Cherepashchuk (\cite{Anatol}) who proposed that such systems should be bright X-ray sources. Following the detection of X-ray emission from massive stars with the {\it EINSTEIN} satellite, and the finding that many of the X-ray brightest massive stars are indeed binaries (Chlebowski \& Garmany \cite{CG}, Pollock \cite{Pollock}), interest in this subject grew rapidly. Over the last decades, enormous progress has been achieved in our understanding of this phenomenon. It is now well established that the signature of wind-wind interactions is not restricted to the sole X-ray domain, but concerns most parts of the electromagnetic spectrum. Therefore, a proper understanding of wind-wind interactions is needed to consistently interpret multi-wavelength observations of massive binaries. 
 
To first order, stellar winds colliding at highly supersonic velocities produce an interaction zone limited by two hydrodynamical shocks. The shape of the contact discontinuity between the two winds is set by ram pressure equilibrium, which is mainly ruled by the so-called wind-momentum ratio ${\cal R} = \left(\frac{\dot{M_1}\,v_{\infty,1}}{\dot{M_2}\,v_{\infty,2}}\right)^{1/2}$ (Stevens \etal\ \cite{SBP}), where $\dot{M}$ and $v_{\infty}$ are the mass-loss rate and the terminal wind velocity, respectively. At the shock fronts, the huge kinetic energy of the incoming flows is converted into heat, leading to a substantial increase of the plasma temperature in the post-shock region. The physical properties of the plasma in the wind interaction zone then depend on the efficiency of radiative cooling (Stevens \etal\ \cite{SBP}). If cooling proceeds very slowly compared to the flow time of the winds, the wind interaction zone is said to be in the adiabatic regime. This regime is mainly encountered in long-period binary systems where the pre-shock densities of the winds are rather low. In this case, the X-ray luminosity is expected to scale with $\frac{\dot{M}^2\,(1 + {\cal R})}{r\,v^{3.2}\,{\cal R}^4}$ where $r$ is the separation between the stars and $v$ the pre-shock velocity (Stevens \etal\ \cite{SBP}). Conversely, if radiative cooling operates very quickly, the wind interaction zone is radiative. This situation frequently occurs in short-period binary systems. In this case, the X-ray luminosity should scale as $\dot{M}\,v^2$.
 
Following the pioneering work of Stevens \etal\ (\cite{SBP}), much insight into the physics of colliding winds has been gained via hydrodynamical simulations. State of the art simulations account for 3-D effects, such as the Coriolis force (Parkin \& Pittard \cite{PP1}, Pittard \& Parkin \cite{PP2}) and clumpy winds (Pittard \cite{Julian}). These models now allow a detailed confrontation with actual observations. 

Generally speaking, the wind interaction will have several consequences that are prone to produce observational signatures. The most obvious examples are the loss of spherical symmetry of the winds, the increase of density and temperature in the post-shock plasma, as well as the acceleration of particles through diffusive shock acceleration. A number of interacting wind systems have now been monitored over a broad range of wavelengths. In addition to high-resolution optical spectroscopy, modern X-ray spectroscopy (with {\it XMM-Newton} and {\it Chandra}) has tremendously contributed to the progress in this field. At the longer wavelength end, {\it Very Long Baseline Interferometer} radio observations allowed us, for the first time, to directly see the wind interaction zone in several systems. 

\section{X-ray emission as a tracer of hot gas in interacting wind systems}
In principle, shock-heated plasma in massive binaries can produce copious amounts of X-rays, exceeding the level of intrinsic X-ray emission from the binary components by a large factor. Indeed, a number of colliding wind binaries have been found to be exceptionally bright in X-rays (e.g.\ WR\,25, Pollock \& Corcoran \cite{PC}). Yet, recent results demonstrated that the vast majority of the massive binaries show a rather `normal' level of X-ray emission (e.g.\ De Becker \etal\ \cite{HD159176}, Naz\'e \cite{2XMM}, Naz\'e \etal\ \cite{Carina}). The reason why large X-ray over-luminosities are restricted to a minority of the massive binaries is still to be understood. Possible explanations include a plasma cooling so efficiently that it emits mostly at longer wavelengths, reductions of the wind pre-shock velocity via radiative braking (Gayley \etal\ \cite{Gayley}), reductions of the mass-loss rates,...

\begin{figure}
\begin{minipage}{5cm}
\includegraphics[width=5cm,angle=-90]{spec000.ps}
\end{minipage}
\hfill
\begin{minipage}{5cm}
\includegraphics[width=5cm,angle=-90]{spec050.ps}
\end{minipage}
\vspace*{2mm}
\begin{minipage}{5cm}
\includegraphics[width=5cm,angle=-90]{spec075.ps}
\end{minipage}
\hfill
\begin{minipage}{5cm}
\includegraphics[width=5cm]{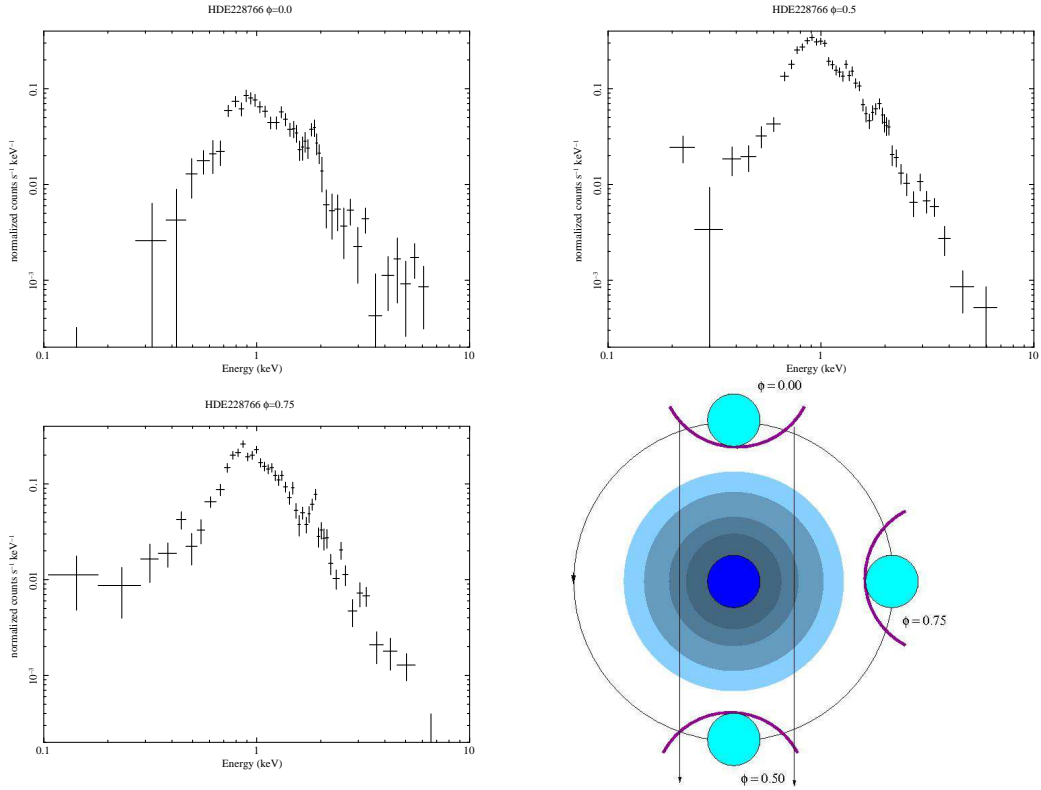}
\end{minipage}
\caption{X-ray spectrum of HDE228766 as a function of orbital phase. Phase 0.0 (upper left) corresponds to the conjunction with the secondary star being in front. Spectra at phases 0.5 and 0.75 are shown by the upper right and lower left panels, respectively. The effect of the dense secondary wind manifests itself through the heavy absorption of the soft X-rays at phase 0.0. To ease comparison, all panels have the same scales. The lower right panel provides a cartoon of the system at the three phases. The observer is located at the bottom of the picture.\label{fig1}}
\end{figure}

In those cases, where a substantial X-ray emission is produced, the observable X-ray flux is often strongly variable with orbital phase. In circular systems, phase-locked variability is due to variations of the occultation and the circumstellar absorption towards the wind interaction zone (see e.g.\ the predictions of the hydrodynamical simulations of Pittard \& Parkin \cite{PP2} for O + O binaries with P$_{\rm orb} \leq 10$\,days). Such a variation is encountered for instance in the 10.7\,day binary HDE\,228766 ($e = 0$). In this system, the primary is a normal O7 star, whilst the secondary is a more evolved Of$^+$-WN8ha transition object (Rauw \etal\ \cite{HDE}). The secondary star is thus expected to have a denser, and more opaque wind. Recent {\it XMM-Newton} observations indeed confirm this picture, with the soft X-ray emission being strongly supressed at conjunction phase with the secondary in front (see Fig.\,\ref{fig1}, Rauw \etal\ in preparation). These differences in absorption can then be used to quantify the mass-loss rates of the winds.

In long-period eccentric systems, one expects to observe considerable variations of the intrinsic level of X-ray emission with the changing separation between the stars. Indeed, if the wind interaction zone is adiabatic, one expects to observe an X-ray flux that varies as $1/r$ where $r$ is the orbital separation. Combining data from an {\it XMM-Newton} and {\it Swift} campaign, Naz\'e \etal\ (\cite{YN2012}) found such a behaviour in the case of Cyg\,OB2 \#9 (O5.5\,I + O3-4\,III, $e=0.71$, P$_{\rm orb} = 860$\,days). 

\begin{figure}[h]
\includegraphics[width=8cm]{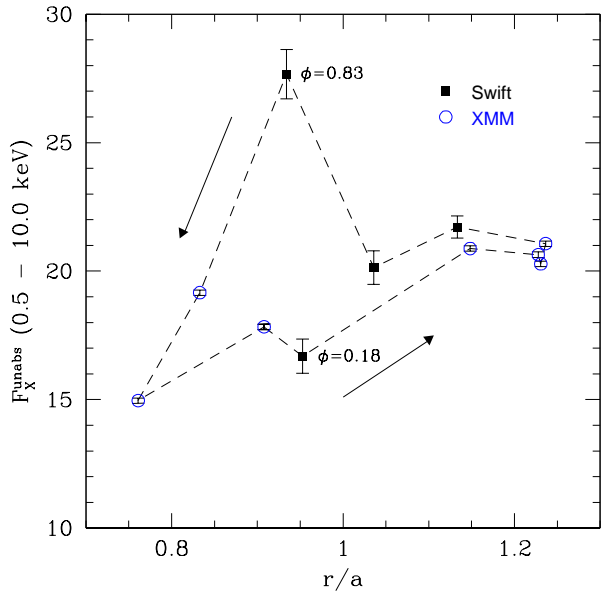}
\caption{X-ray flux of Cyg\,OB2 \#8a, corrected for the interstellar absorption, as a function of orbital separation (see Cazorla \etal\ \cite{Cazorla} for further details).\label{hysteresis}}
\end{figure}

For eccentric O + O systems with intermediate orbital periods, the wind interaction zone remains at least partially radiative and can switch between essentially adiabatic (near apastron) and mainly radiative (near periastron). In such systems, the properties of the wind interaction zone depend on its history, and Pittard \& Parkin (\cite{PP2}) therefore found a strong hysteresis in the X-ray fluxes as a function of orbital separation: the X-ray emission is predicted to be larger and harder at phases prior to periastron passage than at symmetric phases after periastron. A good example of such a behaviour is Cyg\,OB2 \#8a (O6\,I + O5.5\,III, P$_{\rm orb} = 21.9$\,days, $e=0.24$, De Becker \etal\ \cite{Michael8a}). An intense {\it XMM-Newton} and {\it Swift} campaign on this system (Cazorla \etal\ \cite{Cazorla}) revealed a strong hysteresis effect in this system (Fig.\,\ref{hysteresis}), in good agreement with theoretical expectations. 

Whilst theoretical models of interacting winds have successfully reproduced many features seen in the X-ray observations, there remain a number of discrepancies. Probably the most important one concerns the X-ray luminosities. To illustrate this point, let us consider the case of the WN7ha + O binary WR\,22 (P$_{\rm orb}$ = 80.3\,days, $e = 0.56$). Parkin \& Gosset (\cite{PG}) presented 3D hydrodynamical simulations of the wind-wind interaction in this system. The wind of the WN7ha star overwhelms that of its companion and the hydrodynamical simulations indicate that, near periastron, the wind collision collapses onto the surface of the O-star. However, the hydro simulations overestimate the observed flux by more than two orders of magnitude compared to the flux level reported by Gosset \etal\ (\cite{Gosset}), based on  phase-resolved {\it XMM-Newton} observations of WR\,22. Also, the simulations fail to reproduce the observed spectral shape. Parkin \& Gosset (\cite{PG}) suggest that part of the discrepancy could be solved if the wind-wind collision remains attached to the surface of the O-star throughout most of the orbital cycle, and the mass-loss rates of both stars are reduced whilst simultaneously increasing the wind momentum ratio in favour of the WN7ha star. 

In some cases, the failure to reproduce the correct X-ray luminosity might be related to the fate of clumps. Indeed, it is now well established that stellar winds of massive stars are clumpy. In this context, Zhekov (\cite{Zhekov}) analysed the X-ray emission from a sample of seven short-period WR + O binaries. The wind interaction zones in these systems were expected to be radiative. However, Zhekov (\cite{Zhekov}) found no correlation between the X-ray and wind luminosities. Instead, the X-ray luminosities follow the scaling law expected for adiabatic wind interaction zones, thus suggesting that the wind interactions are in fact adiabatic. This can only be the case if only part of the wind (i.e.\ the homogeneous, inter-clump component) contributes to the X-ray emission of the wind-wind collision. In this scenario, the clumps would pass freely through the interaction zone, without being destroyed. This contrasts with the results of the simulations of Pittard (\cite{Julian}) who found that the clumps dissolve as they cross the shock. However, the latter simulations were done for a long-period (wide) binary system, and simulations of shorter period systems are needed to further understand the implications of the results of Zhekov (\cite{Zhekov}).
 
\section{Evidence for wind interactions in the optical domain}
The loss of spherical symmetry of the winds as well as the overdensity of the post-shock plasma affect the line profiles of optical, UV and IR emission lines that form in the wind (e.g.\ Stevens \cite{IRS}). If the shocked material cools efficiently, the shock region collapses, resulting in very large overdensities of the material in the wind interaction zone with respect to the ambient winds. Therefore, emission lines that are produced through recombination (and have thus strengths proportional to $\rho^2$) form partially in the wind interaction zone. The phase-dependence of the line profiles can be used to constrain the properties of the wind interaction, as well as the orbital inclination via a model originally deviced by L\"uhrs (\cite{Luehrs}).
 
The orbital motion introduces a curvature of the wind collision region at some distance from the stars. This feature can be observed directly in the so-called pinwheel nebulae where dust is formed in the outer parts of a highly curved interaction zone of a massive binary containing a carbon-rich WC-star (Tuthill \cite{Tuthill}). Forming dust in the hostile environment of a massive binary is extremely difficult: close to the stars, the harsh UV radiation field is too strong, and at larger distances, where the UV field is diluted, the density of the wind has dropped to too low a level for dust formation. Only a wind-wind interaction with efficient cooling can provide the necessary increase in density that allows dust to form. In some systems this happens only around specific orbital phases. This is the case for instance in  WR\,140 (WC7 + O5.5, P$_{\rm orb} = 7.94$\,years, $e = 0.88$, Williams \cite{Peredur}) where episodes of strong IR emission, attributed to dust formation, occur at periastron passage. In some other systems, dust is seen persistently. An example is WR\,70 (WC9 + B0\,I, Williams \etal\ \cite{WR70}) which displays a persistent, but variable circumstellar dust emission. These authors found a best-fit period to the IR variations of 1030 days (2.82\,years), which could thus reflect the orbital period, although the behaviour is not strictly regular. In WR\,70, the fraction of carbon atoms of the WC9 wind going into the dust production varies between about 11 and 46\% (Williams \etal\ \cite{WR70}). 

In eccentric binary systems, the curvature of the wind interaction zone changes with orbital phase, being minimum at apastron and maximum at periastron passage. An extreme example of such a situation is found in $\eta$\,Car, a highly eccentric binary ($e \sim 0.9$) with an orbital period of 5.54\,years. The primary star is an LBV with a mass-loss rate of $10^{-3}$\,M$_{\odot}$\,yr$^{-1}$ and a low wind velocity of 500 -- 600\,km\,s$^{-1}$. The secondary is unseen, but is likely a mid-O supergiant with $\dot{M} \sim 10^{-5}$\,M$_{\odot}$\,yr$^{-1}$ and $v_{\infty} \sim 3000$\,km\,s$^{-1}$. The wind-wind interaction results in an extended cavity carved by the secondary wind in the wind of the LBV primary component. This cavity is bordered by high-density shells of the shocked primary wind that form extended spirals. Okazaki \etal\ (\cite{Okazaki}) used smooth particle simulations to reproduce $\eta$\,Car's X-ray light curve as observed with {\it RXTE}. These simulations were subsequently used by Madura \etal\ (\cite{Madura}) to successfully reproduce the phase dependence of the spatial and radial velocity distribution of the [Fe\,{\sc iii}] $\lambda$\,4659 line as observed with {\it HST}/STIS. This high-ionization forbidden line emission arises in the extended primary wind and wind-wind collision regions that are photoionized by the hot secondary component, and have the right density and temperature for producing the forbidden line. The shape and extent of this emission region change with orbital phase (Gull \etal\ \cite{Gull}), the photoionization region being most compact and the [Fe\,{\sc iii}] emission being suppressed near periastron passage (Madura \etal\ \cite{Madura}). These models allowed Madura \etal\ (\cite{Madura}) to show that the orbital axis is closely aligned with the symmetry axis of the Homunculus nebula. 

When the components of $\eta$\,Car are approaching periastron, the strength of the He\,{\sc ii} $\lambda$\,4686 emission line rises suddenly up to EW $\sim -2.5$\,\AA. Around periastron, the equivalent width sharply drops to zero, before rising again and then decline (Teodoro \etal\ \cite{Teodoro}). The most likely place where this line is formed is the shocked primary wind, indicating that the sharp decline is likely a result of an occultation of the emission region by the thick primary wind. The intrinsic luminosity of this line is up to 250 -- 300\,L$_{\odot}$, larger than the X-ray luminosity in the 2 -- 10\,keV band (Teodoro \etal\ \cite{Teodoro}). The variations of the He\,{\sc ii} $\lambda$\,4686 emission line are delayed by about 16.5\,days with respect to the variations of the X-ray flux. This is attributed to the flow time from the apex of the wind-wind collision zone (where most of the hard X-rays are emitted) to the He\,{\sc ii} $\lambda$\,4686 emission region. Such a delay is expected as a result of the time needed for the post-shock plasma to cool down sufficiently for the formation of the optical recombination line.

\section{Relativistic particles in colliding wind binaries}
Diffusive shock acceleration in hydrodynamical shocks can accelerate particles up to relativistic energies through the first order Fermi mechanism (Bell \cite{Bell1}, \cite{Bell2}). When relativistic electrons interact with a magnetic field, they produce synchrotron radio emission (Eichler \& Usov \cite{EU}, Pittard \etal\ \cite{Pittard2006}, Pittard \& Dougherty \cite{PD}). 

From an observational point of view, a subset of massive stars display such a non-thermal radio emission associated with a wind-wind interaction (De Becker \& Raucq \cite{catalogNT}, and references therein). This non-thermal radio emission is often variable as a result of changing line-of-sight optical depth and, in eccentric systems, of changing intrinsic emission (White \& Becker \cite{WB}, Blomme \etal\ \cite{Blomme2010}, \cite{Blomme2012}). {\it VLBI} observations allowed to resolve the emitting region of the non-thermal radio emission in several systems, revealing an arc-like morphology, consistent with a wind-wind interaction (Dougherty \etal\ \cite{Dougherty}, Ortiz-Le\'on et al.\ \cite{OL}). 

The presence of relativistic electrons, combined with the enormous supply of UV photons by the binary components, should result in a strong inverse Compton scattering emission in the hard X-rays and soft $\gamma$-rays (Pollock \cite{Pollock2}, Chen \& White \cite{CW}, Pittard \& Dougherty \cite{PD}). Such an emission was indeed detected in $\eta$\,Carinae (Leyder \etal\ \cite{Leyder}) and WR\,140 (Sugawara \etal\ \cite{Sugawara}). In addition, $\gamma$-ray emission associated with $\pi^0$ decay was also detected in $\eta$\,Car (Farnier \etal\ \cite{Farnier}, Reitberger \etal\ \cite{Reitberger}), showing that particle acceleration in interacting wind regions is not restricted to electrons. 

An open issue in this context, is why the non-thermal radio emission is only seen for a subset of the known interacting wind systems. 

\section{Conclusions}
Interacting winds are inherent to massive binary systems. They produce a number of exciting properties and much progress has been achieved over recent years. Further insight into the dynamics of the phenomenon will be made with future instrumentations, such as the integral-field high-resolution X-ray spectrometer aboard the {\it Athena+} mission, currently proposed to ESA (Sciortino \etal\ \cite{Athena}).


\end{document}